\documentclass[twocolumn,prb,showpacs]{revtex4}
\usepackage{graphicx}

\begin{document}

\title{Self-focusing magnetostatic beams in thin magnetic films}

\author{Ramaz Khomeriki} 
\thanks{E-mail address: khomeriki@hotmail.com}

\affiliation{Department of Physics, Tbilisi
State University, Chavchavadze ave.\ 3, Tbilisi 0128, Republic of Georgia}

\begin{abstract}
The possibility of generation of stable self-focusing beams in in-plane magnetized thin magnetic films is considered and theoretical conditions for the existence of such localized solutions are discussed. It is shown that for the definite direction between static magnetizing field and preferential direction of radiation from microwave antenna the problem reduces to the one-dimensional nonlinear Schr\"odinger equation. For such angles it is possible to generate stable self-focusing beams. Particular values of beam width and propagation angles versus magnitude of magnetizing field are calculated in order to suggest the realistic experimental setup for the observation of discovered effect. 

\end{abstract}

\pacs{85.70.Ge; 75.30Ds; 76.50.+g}

\maketitle

\section{Introduction}

Observations of magnetostatic solitons in thin magnetic films together with experiments in nonlinear optics are major testing grounds for the nowadays advances in nonlinear physics. In full accordance with theoretical predictions \cite{zvezdin} magnetostatic bright envelope solitons have been observed in both in-plane \cite{boyle,chen1,kalinikos1,xia} and perpendicularly magnetized \cite{f1,f2,f3} quasi-one-dimensional yttrium-iron garnet thin films (magnetic waveguides). On the other hand, as expected, dark surface wave magnetostatic solitons have been observed only in in-plane magnetized films \cite{chen2,slavin,kalinikos}. Moreover, in full analogy with light bullets in nonlinear optics \cite{o1,o2} spin-wave metastable bullets have been found in wide magnetic films \cite{zalit1,zalit2}. The only difference between nonlinear processes in magnetic films and optical devices is that self-focusing magmetostatic beams are unstable at relatively long distances unlike their optical analogies. In particular, in case of magnetic films the focusing into one spatial point takes place in stationary regime \cite{ref}. This is explained by the fact that longitudinal dispersion could not be neglected in magnetic films. In the present paper it is shown that even that gap could be filled considering in-plane magnetized films where carrier wave vector is not either parallel or perpendicular to the static magnetic field direction. The conditions when stationary and stable self-focusing magnetostatic beams can be observed in wide magnetic films are found. 

Although the linearized spin-wave solutions are well known for arbitrary directions between wave vector and magnetizing field \cite{damon,hurben}, the nonlinear situation has been studied only for the cases when carrier wave vector is either parallel or perpendicular to the magnetizing field direction \cite{zvezdin,boyle,chen1,kalinikos1,xia,f1,f2,f3,chen2,slavin,kalinikos,zalit1,zalit2,ref} (one exception is Ref. \cite{referee} where magnetized field is tilted from the film normal in order to control the nonlinear coefficient, but this study does not pertain to the present consideration). Only very recently the nonlinear effects characterizing general case has been investigated \cite{ramazlasha} and soliton solutions have been found for the angles between wave vector and magnetic field other than 0 or 90 degrees. However, such solutions are stable only in quasi-one-dimensional case and they become unstable considering wide samples. As it will be shown below in two dimensional case only self-focusing beam solutions are stable. Note that bullet like solutions are metastable and they decay after either edge reflection or interaction with other moving localization \cite{zalit1,zalit2} (metastability takes place due to the compensation of instability by dissipation).

It should be especially mentioned that the wave processes are easily accessible from the surface by variety of the methods such as inductive probes \cite{exp1}, thermo-optical methods \cite{exp2} and recently developed method of space and time resolved Brillouin light scattering \cite{exp3}. Therefore it does not seem problematic to detect nonlinear localizations predicted in this paper.

\section{Problem geometry and reduction to 1D NLS}

For arbitrary direction of the carrier wave vector $\vec k$ with respect to a static magnetic field $\vec H$ the problem of nonlinear wave process (magnetostatic and Landau-Lifshitz equations) in in-plane magnetized thin film reduces to the following nonlinear equation for the wave envelope $u$ (see e.g. Ref. \cite{ramazlasha}):
\begin{eqnarray}
i\left(\frac{\partial u}{\partial t}+v_y \frac{\partial u}{\partial y}+ v_z \frac{\partial u}{\partial z}\right)+\frac{\omega''_{yy}}{2}
\frac{\partial^2 u}{\partial y^2}+\frac{\omega''_{zz}}{2}
\frac{\partial^2 u}{\partial z^2}+ \nonumber \\
\omega''_{yz}
\frac{\partial^2 u}{\partial y\partial z}-N|u|^2u= -i\omega_r u, \label{1}
\end{eqnarray}
where $\vec r$ stands for the radius vector lying in the sample plane $yz$, $x$ is a coordinate along the direction perpendicular to the film, and $z$ is a direction of static magnetic field; $v_y$ and $v_z$ are the components of group velocity and they could be calculated from the linearized dispersion relation (but see below) $\vec v\equiv \partial \omega/\partial \vec k$; $\omega$ and $\vec k$ are carrier frequency and wave vector of magnetostatic spin wave; $\omega''_{\beta\gamma}\equiv\partial^2\omega/\partial k_\beta\partial k_\gamma$ (indexes $\beta$ and $\gamma$ take the values $y$ and $z$). $u$ is a complex envelope of relative magnetization vector $\vec m=\vec M/M_0$ ($M_0$ is a static magnetization along $z$ coinciding with the direction of magnetizing field):
\begin{eqnarray}
m_x&=&u~\frac{ik_x\omega_H-k_y\omega_0}{k\sqrt{\omega_H^2+\omega_0^2}} ~e^{i(\omega t-\vec{k}\vec{r})};  \nonumber \\ &&
m_y=u~\frac{k_x\omega_0+ik_y\omega_H}{k\sqrt{\omega_H^2+\omega_0^2}}~e^{i(\omega t-\vec{k}\vec{r})}. \label{ref}
\end{eqnarray}
Here $\omega_H=gH$, $\omega_M=4\pi gM_0$ and $\omega_0\equiv\omega(k=0)=\sqrt{\omega_H(\omega_H+\omega_M)}$ ($g$ is modulus of gyromagnetic ratio for electrons). The above expressions (\ref{ref}) have been derived \cite{ramazlasha} in the limit $kd\ll 1$ ($d$ is a film thickness) and this condition will be used further in this paper in order to simplify analytical calculations. Then the dispersion relation could be expressed as expansion over small parameter $kd$ and keeping only the terms up to the second 
order of this parameter the following expression is obtained \cite{ramazlasha}: 
\begin{eqnarray}
\omega=\omega_0+\frac{\omega_M}{4\omega_0}\frac{d}{k}\left(\omega_Mk_y^2-
\omega_Hk_z^2\right)- ~~~~~~~~~ \\
\frac{\omega_M^2}{32\omega_0^3}\frac{d^2}{k^2}
\left(\omega_Mk_y^2- 
\omega_Hk_z^2\right)^2+\frac{\omega_M}{4\omega_0}d^2\left(\omega_H\frac{k_z^2}{3}-\omega_Mk_y^2\right).  \label{2} \nonumber
\end{eqnarray}
The phenomenological dissipation parameter $\omega_r$ in (\ref{1}) is weak, but it plays important role in stabilization of localized solutions (see e.g. Ref. \cite{ref}) and finally, the nonlinear coefficient $N$ in case of in-plane magnetized films is always negative and reads as follows \cite{zvezdin} $N=-\omega_H\omega_M/4\omega_0$. 
\begin{figure}[t]
\begin{center}\leavevmode
\includegraphics[width=\linewidth]{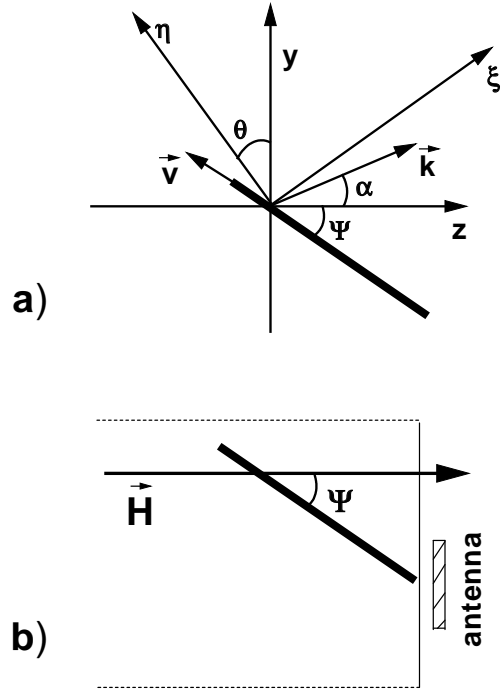}
\vspace{-2.5cm}
\caption{Geometry of the problem in case of stationary self-focusing beam. a) Orientations of wave vector $\vec k$, static magnetic field $\vec H$, group velocity $\vec v$ and diagonalizing frame of references $\eta\xi$. Thin line indicates the direction of the beam. b) The possible experimental setup for observation of stable magnetostatic self-focusing beam. $\Psi$ indicates an angle between static magnetizing field and preferential radiation direction of antenna (see also Figs. 5a and 6b in Ref. \cite{ref}).}
\label{run0}\end{center}\end{figure}

In order to vanish the nondiagonal term with coefficient $\omega''_{yz}$ in Eq. (\ref{1}) a new frame of references $\eta\xi$ should be introduced (see Fig. \ref{run0}a). Let us rotate the frame of references $yz$ by the angle $\vartheta$ defined from the following relation:
$\tan(2\vartheta)=2\omega''_{yz}/(\omega''_{zz}-\omega''_{yy})$.
Then from (\ref{1}) (2+1) dimensional (two spatial and one temporal dimensions) nonlinear Schr\"odinger (NLS) equation is obtained:
\begin{eqnarray}
i\left(\frac{\partial u}{\partial t}+v_\xi\frac{\partial u}{\partial \xi}+
v_\eta\frac{\partial u}{\partial \eta}\right)&+&\frac{R}{2}
\frac{\partial^2 u}{\partial \xi^2}+\frac{S}{2}
\frac{\partial^2 u}{\partial \eta^2} \nonumber \\ && -  
N|u|^2u= -i\omega_r u, \label{3}
\end{eqnarray}
which reduces to the standard form (without first spatial derivatives) after introducing the following coordinate transform (moving frame) $\eta\rightarrow\eta- v_\eta t$ and $\xi\rightarrow\xi- v_\xi t$. In Eq. (\ref{3}) coefficients
$R$ and $S$ are dispersion and diffraction coefficients, respectively, explicit form of which are given in Ref. \cite{ramazlasha} and 
$v_\eta$ and $v_{\xi}$ are group velocity components with respect to new reference frame $\eta\xi$. Usually in isotropic systems transversal component of group velocity $v_\eta$ is equal to zero (the same happens in case of nonlinear magnetostatic waves when carrier wave vector is either parallel or perpendicular to a static magnetic field \cite{zvezdin,boyle}). But in general for anisotropic systems $v_\eta$ is not equal to zero. 

As well known (2+1) NLS equation does not permit \cite{zakharov,kivshar} stable localized solutions irrespective to the relative sign of the coefficients $S$, $R$ and $N$. Only the metastable bullet like localizations appear \cite{o1,o2,zalit1,zalit2} due to the compensation of wave instability by the dissipation. However, in restricted geometries (waveguides) transverse instabilities do not develop and diffraction term could be neglected allowing thus reduction to (1+1) NLS equation. Such geometries have been used in order to observe solitons in optical fibers and magnetic film waveguides. 

Another possibility to see localized solution is the absence of dispersion term $R=0$. Such a situation is realized in nonlinear optics where dispersion is negligible in comparison with the diffraction. In this case spatial soliton solution (self-focusing beam) is stable \cite{cigni}. However, in case of magnetic films dispersion coefficient could not be neglected and beam like solutions have not been observed in the experimental conditions considered till now. In the present paper it is suggested the experimental setup where the angle $\alpha$ between carrier wave vector and magnetizing field gets the value for which dispersion coefficient is nearly zero ($R\simeq 0$). As far as all the coefficients of (2+1) NLS equation (\ref{3}) are defined from the dispersion relation (\ref{2}), they could be expressed as functions of wave number $k$ and the angle $\alpha$. Particularly, in the limit of small $k$ ($kd\rightarrow 0$) dispersion coefficient $R$ does not depend on $k$. Therefore the problem is reduced to finding such $\alpha$ which makes $R$ equal to zero for given static field and sample parameters. As numerical simulations show, for each magnitude of static magnetic field it is possible to find such an angle.

Actually it is not necessary to have such $\alpha$-s for which $R$ is exactly zero. One can neglect the role of dispersion in formation of nonlinear wave if the following inequality holds $R\ll\omega_r/k^2$, i.e. the wave dissipates faster than dispersion effects take place. Similarly one can neglect higher order dispersion terms in comparison with dissipation as far as they are proportional to the factor $(kd)^3$ ($kd\ll 1$ in this paper). At the same time the diffraction term in my calculations is much larger than dissipation one. Thus only the diffraction and nonlinearity determine the dynamics of nonlinear wave and reduction to (1+1) NLS equation is justified.

\section{Stationary self focusing beam solution}

Considering standard stationary situation $\partial/\partial t=0$ and using coordinate transform $\eta\rightarrow\eta-(v_\eta/v_{\xi})\xi$ in case of close to zero dispersion ($R\simeq 0$) Eq. (\ref{3}) reduces to (1+1) NLS equation ($\xi$ plays a role of the time) with stable spatial soliton solutions. For instance one soliton solution could be presented analytically as follows:
\begin{equation}
|u|=|u|_{max}sech\left\{\frac{\eta-(v_\eta/v_{\xi})\xi}{\Lambda}\right\} \label{4}
\end{equation}
corresponding to the self-focusing beam along the direction of group velocity $\vec v$. Here beam width $\Lambda$ is defined as follows:
\begin{equation}
\Lambda=\left|\frac{S}{N}\right|^{1/2}\frac{1}{|u|_{max}}. \label{5}
\end{equation}
Note that the amplitude $|u|_{max}$ decays and beam width increases with distance taking into account the weak dissipation effects.

\begin{figure}[t]
\begin{center}\leavevmode
\includegraphics[width=0.9\linewidth]{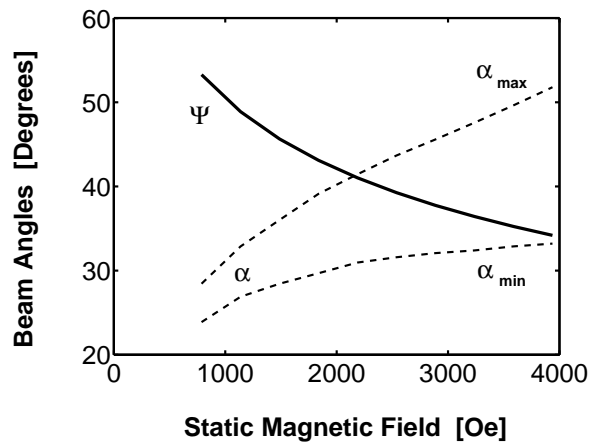}
\caption{Range of angles for which self-focusing beam regime could be realized. $\alpha$ (dashed borders) stands for the angle range between carrier wavevector and static magnetic field, while $\Psi$ (solid line) is a  corresponding angle between direction of group velocity and static magnetic field.}
\label{run}\end{center}\end{figure}

Knowing angles $\alpha$ for which the dispersion effect could be neglected, it is easy to calculate diffraction coefficient $S$ and group velocity $\vec v$. Then the angle $\Psi$ between $\vec v$ and axis $z$ for that particular values of angles $\alpha$ could be found. The observation of stationary self-focusing is possible only for the mentioned radiation direction. In Fig. \ref{run} the dependences of angles $\alpha$ and $\Psi$ versus static magnetic field are presented. 

\begin{figure}[t]
\begin{center}\leavevmode
\includegraphics[width=0.9\linewidth]{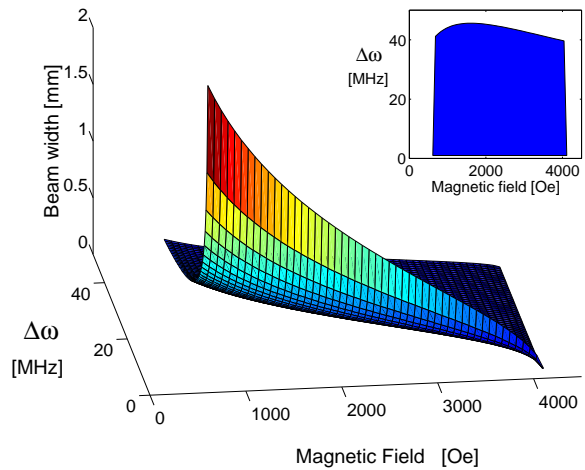}
\caption{Self focusing beam width versus detuning of the carrier frequency $\Delta\omega=\omega_0-\omega$ and static magnetic field. Inset shows the projection of the plot on the horizontal plane where the filled area indicates considered range. static magnetic field is restricted by the boundaries $0.3\omega_M<\omega_H<2.37\omega_M$ (see the text), while the upper limit corresponds to the condition $kd<0.1$. The following parameters are used for the calculations: relative amplitude of the beam $|u|_{max}=0.1$; film thickness $d=5\mu m$ and demagnetizing field $H_M=1750$ Oe.}
\label{run3d}\end{center}\end{figure}

\section{Discussion of possible experimental setup}

The following values for the film parameters are used in the calculations: film thickness $d=5\mu m$; demagnetizing field $H_M=\omega_M/g=1750$ Oe and dissipation parameter is taken $\omega_r=5\cdot 10^6$s$^{-1}$ as in Ref. \cite{ref}. As it is discussed above the beam self-focusing process is realized if dispersion effects could be neglected, i.e. if the following inequality is satisfied $R\ll \omega_r/k^2$. The calculations are made for carrier wave number $k=50$cm$^{-1}$. In Fig. \ref{run} the angle range between $\vec k$ and static magnetic field is presented for which the above inequality holds (dispersion effects could be neglected) and, besides that, the range of angles $\alpha$ corresponds to the single direction of group velocity. As seen for the values of static magnetic field $H_0>2500 Oe$ almost the same direction of group velocity corresponds to the wide range of validity of beam generation regime. That direction must coincide with a preferential direction of magnetostatic wave radiation from short antenna or point like source. The ways to vary experimentally the preferential direction of radiation in linear regime has been suggested in Ref. \cite{ref} and now we suggest here to use the same method in nonlinear regime in order to observe self-focusing beams. Thus the antenna should be oriented by such a way that the angle between its preferential direction of radiation and static magnetic field coincides with the derived angles for beam group velocity (angles $\Psi$ in Figs. \ref{run0} and \ref{run}).  It should be mentioned that as calculations show in the considered limit $kd\rightarrow 0$ the required angles between radiation direction and static magnetic field do not depend on $|k|$ and consequently on the carrier frequency $\omega$ of the excitation (note that this happens when dispersion coefficient $R$ is negligible). While the diffraction coefficient $S$ and as a result beams width is inverse proportional to $kd$. 

In Fig. \ref{run3d} three dimensional plot of beam width $\Lambda$ versus detuning of carrier frequency $\Delta\omega=\omega_0-\omega$ and magnitude of static magnetic field is presented. Magnetic field varies within the boundaries $0.3\omega_M<\omega_H<2.37\omega_M$ where the lower boundary appears from the requirement that three magnon processes should not take place \cite{zvezdin} (otherwise the localizations will decay rapidly), while above a upper limit the diffraction coefficient $S$ becomes negative and consequently according to the Lighthill criterion \cite{lighthill} self-focusing process does not take place.

\section{Conclusions}

Concluding it could be stated that the conditions for the observation of stationary self-focusing beams in magnetic films are found. It is suggested that such localizations could be observed along preferential direction of antenna's radiation. 

{\bf Acknowledgement:} The research described in this publication was made possible in part by Award No. GP2-2311-TB-02 of the U.S. Civilian Research \& Development Foundation for the Independent States of the Former Soviet Union (CRDF) and NATO reintegration grant No. FEL.RIG.980767.


\begin{thebibliography}{00}

\bibitem{zvezdin} A.K.  Zvezdin, A.F.  Popkov, Zh. Eksp. Theor. Phys., {\bf 84}, 606, (1983).
\bibitem{boyle} J.W.  Boyle, S.A.  Nikitov, A.D.  Boardman, J.G.  Booth, K.
Booth, Phys.Rev.B., {\bf 53}, 12173, (1996).
\bibitem{chen1} M.  Chen, M.A.  Tsankov,
J.M.  Nash, C.E.  Patton, Phys.Rev.B., {\bf 49}, 12773, (1994).
\bibitem{kalinikos1} N.G.  Kovshikov, B.A.  Kalinikos, C.E.  Patton, E.S.
Wright, J.M.  Nash, Phys.Rev.B., {\bf 54}, 15210, (1996).
\bibitem{xia} H.  Xia, P.  Kabos, C.E.  Patton, H.E.  Ensle, Phys.Rev.B., 
{\bf
55}, 15018, (1997).
\bibitem{f1} B.A. Kalinikos, N.G. Kovshikov, A.N. Slavin, Zh. Eksp. Teor. Fiz. {\bf 94}, 159, (1988).
\bibitem{f2} P. De Gasperis, R. Marcelli, G. Miccoli, Phys. Rev. Lett. {\bf 59}, 481, (1987).
\bibitem{f3} B.A. Kalinikos, N.G. Kovshikov, A.N. Slavin, Phys. Rev. B {\bf 42}, 8658, (1990)
\bibitem{chen2} M. Chen, M.A. Tsankov, J.M. Nash, C.E. Patton, Phys. Rev. 
Lett.,
{\bf 70}, 1707, (1993).
\bibitem{slavin} A.N. Slavin, Yu.S. Kivshar, E.A. Ostrovskaya, H. Benner, 
Phys. Rev. Lett.,
{\bf 82}, 2583, (1999).
\bibitem{kalinikos} B.A. Kalinikos, M.M. Scott, C.E. Patton, Phys. Rev. 
Lett., {\bf 84}, 4697, (2000).
\bibitem{o1} Y. Silberberg, Opt. Lett., {\bf 15}, 1282, (1990).
\bibitem{o2} X. Liu, L.J. Qian, F.W. Wise, Phys. Rev. Lett., {\bf 82}, 4631, (1999).
\bibitem{zalit1} M.  Bauer, O.  B\"uttner, S.O.  Demokritov, B.  
Hillebrands, V. Grimalsky, Yu.  Rapoport, A.N.  Slavin, Phys.  Rev.  Lett., {\bf 81}, 3769, (1998).
\bibitem{zalit2} O.  B\"uttner, M.  Bauer, S.O.  Demokritov, B.
Hillebrands, M.P.  Kostilev, B.A.  Kalinikos, A.N.  Slavin, Phys.  Rev.  
Lett., {\bf 82}, 4320, (1999).
\bibitem{ref} O.  B\"uttner, M.  Bauer, S.O.  Demokritov, B.
Hillebrands, Yu.S. Kivshar, V. Grimalsky, Yu. Rapoport, A.N.  Slavin, 
Phys.  Rev. B, {\bf 61}, 11576, (2000).
\bibitem{cigni} R. K. Dodd , J. C. Eilbeck, J. D. Gibbon, H. C. Morris, {\it Solitons and nonlinear wave equations}, London, Academic Press, (1982), Chapt. 8. 
\bibitem{damon} R.W. Damon, J.R. Eshbach, J. Phys. Chem. Solids, {\bf 19},
308, (1961).
\bibitem{hurben} M.J. Hurben, C.E. Patton, J. Magn. Magn. Mater., {\bf 139},
263, (1995).
\bibitem{referee} B.A. Kalinikos, N.G. kovshikov, A.N. Slavin, IEEE Transactions on Magnetics, {\bf 26}, 1477, (1990).
\bibitem{ramazlasha} R. Khomeriki, L. Tkeshelashvili, Phys. Rev. B, {\bf 65}, 134415, (2002).
\bibitem{exp1} N.P. Vlannes, J. Appl. Phys., {\bf 61}, 416, (1987).
\bibitem{exp2} O.V. Geissau, U. Netzelmann, S.M. Rezende, J. Pelzl, IEEE Trans. Magn., {\bf 26}, 1471, (1990).
\bibitem{exp3} B. Hillebrands, Rev. Sci. Insrum., {\bf 70}, 1589, (1999).
\bibitem{zakharov} V.E. Zakharov, A.M. Rubenchik, Zh. Eksp. Teor. Fiz., {\bf 
65},
997, (1973); [Sov. Phys. JETP, {\bf 38}, 494, (1974)].
\bibitem{kivshar} Yu.S. Kivshar, D.E. Pelinovsky, Physics Reports, {\bf 
331}, 117, (2000).
\bibitem{lighthill} M.J. Lighthill, J. Inst. Math. Appl. {\bf 1}, 269, (1965).



\end{thebibliography}
\end{document}